\documentclass{iau}
\usepackage{graphicx}

\def\ga{\mathrel{\hbox{\rlap{\hbox{\lower4pt\hbox{$\sim$}}}\hbox{$>$}}}}
\newcommand{\Msolar}{\mbox{\,$\rm M_{\odot}$}}	      % solar mass
\newcommand{\Rsolar}{\mbox{\,$\rm R_{\odot}$}}	      % solar radius
\newcommand{\vsini}{$\!${\em v\,}sin{\em i} }

\newcommand{\sups}[1]{$^{\rm #1}$}

\title[Starspots on Young Solar-Type Stars] %% give here short title %%
{Starspots on Young Solar-Type Stars}

\author[Brown et al.]   %% give here short author list %%
{Carolyn Brown,$^1$ Brad Carter,$^1$ Stephen Marsden$^1$ \and Ian Waite$^1$}

\affiliation{$^1$Computational Engineering and Science Research Centre, University of Southern Queensland, \\Toowoomba, Australia \\ email: {\tt carolyn.brown@usq.edu.au}}

\pubyear{2013}
\volume{302}  %% insert here IAU Symposium No.
\pagerange{--}
\jname{Magnetic Fields Throughout Stellar Evolution}
\editors{P. Petit, M.M. Jardine \& H.C. Spruit, eds.}
\begin{document}

\maketitle

\begin{abstract}
Doppler Imaging of starspots on young solar analogues is a way to investigate the early history of solar magnetic activity by proxy. Doppler images of young G-dwarfs have yielded the presence of large polar spots, extending to moderate latitudes, along with measurements of the surface differential rotation. The differential rotation measurement for one star (RX J0850.1-7554) suggests it is possibly the first example of a young G-type dwarf whose surface rotates as almost a solid body, in marked contrast to the differential rotation of other rapidly rotating young G-dwarfs and the present-day Sun. Overall, our Doppler imaging results show that the young Sun possessed a fundamentally different dynamo to today.
%% add here a maximum of 10 keywords, to be taken form the file <Keywords.txt>
\end{abstract}

\firstsection % if your document starts with a section,
              % remove some space above using this command.
\section{Introduction}

Active young solar-type stars provide proxies for studying the early evolution of the Sun's dynamo and activity. Here we present preliminary Doppler Imaging (DI)  results and differential rotation measurements for three rapidly rotating (\vsini $>$ 20 km s\sups{-1}) active young G-type stars with similar radii (1.09 - 1.22 \Rsolar) and masses (1.15 - 1.20 \Msolar) to the Sun. These stars were chosen as they also represent different stages of early solar evolution, with 17 Myr old RX J0850.1-7554 being a Pre Main Sequence star, 25 Myr old LQ Lup a post-T Tauri star, and 35 Myr old R58 being on the Zero Age Main Sequence.

\section{Methodology}

The observations presented here were all taken with the same instrumentation setup at the Anglo-Australian Telescope utilising the high resolution {\'e}chelle spectrograph, UCLES, to output the spectroscopic data on the CCD. Least-Squares Deconvolution (LSD, \cite[Donati et al. 1997]{Donati_etal97}) was utilised to increase the signal-to-noise to a level where DI could be carried out for the selected targets using the maximum entropy image reconstruction code described in \cite[Brown et al. (1991)]{Brown_etal91}. The surface differential rotation of the stars was determined using the the $\chi^{2}$-minimisation technique of \cite[Petit et al. (2002)]{PetitP_2002}.

\section{Results and Discussion}

\begin{figure}[t]
\begin{center}
\includegraphics[width=3.2cm]{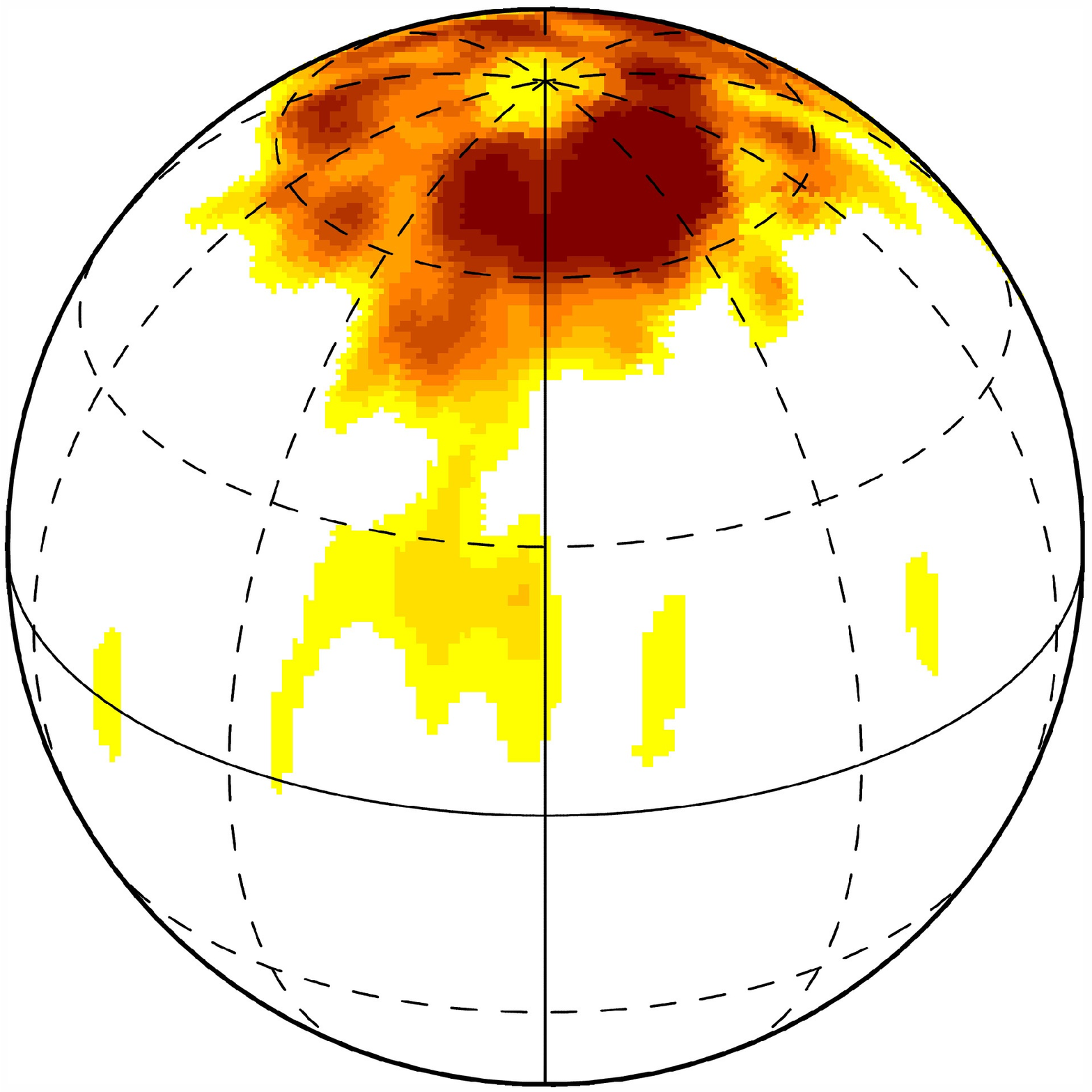}
\includegraphics[width=3.2cm]{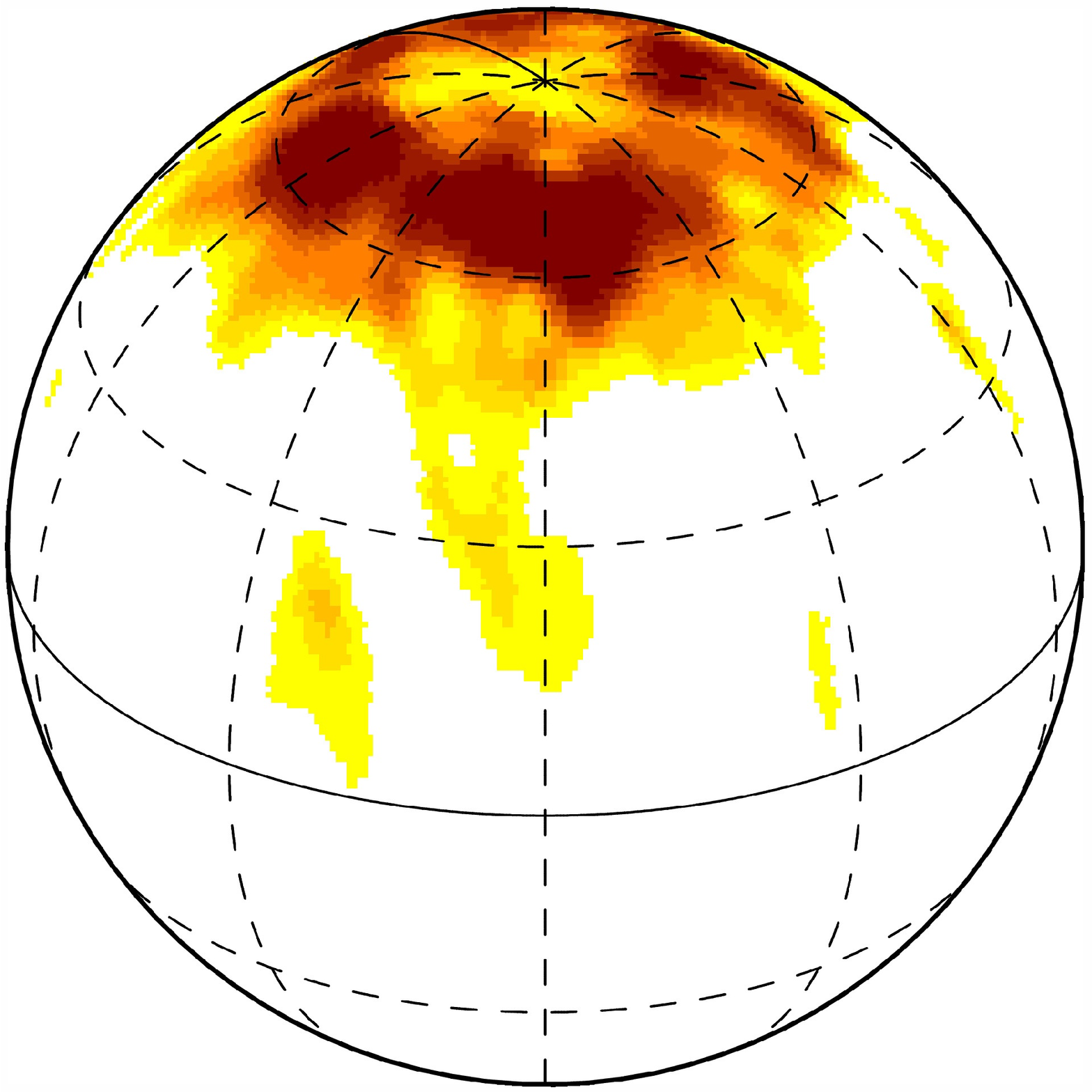}\\
\includegraphics[width=3.2cm]{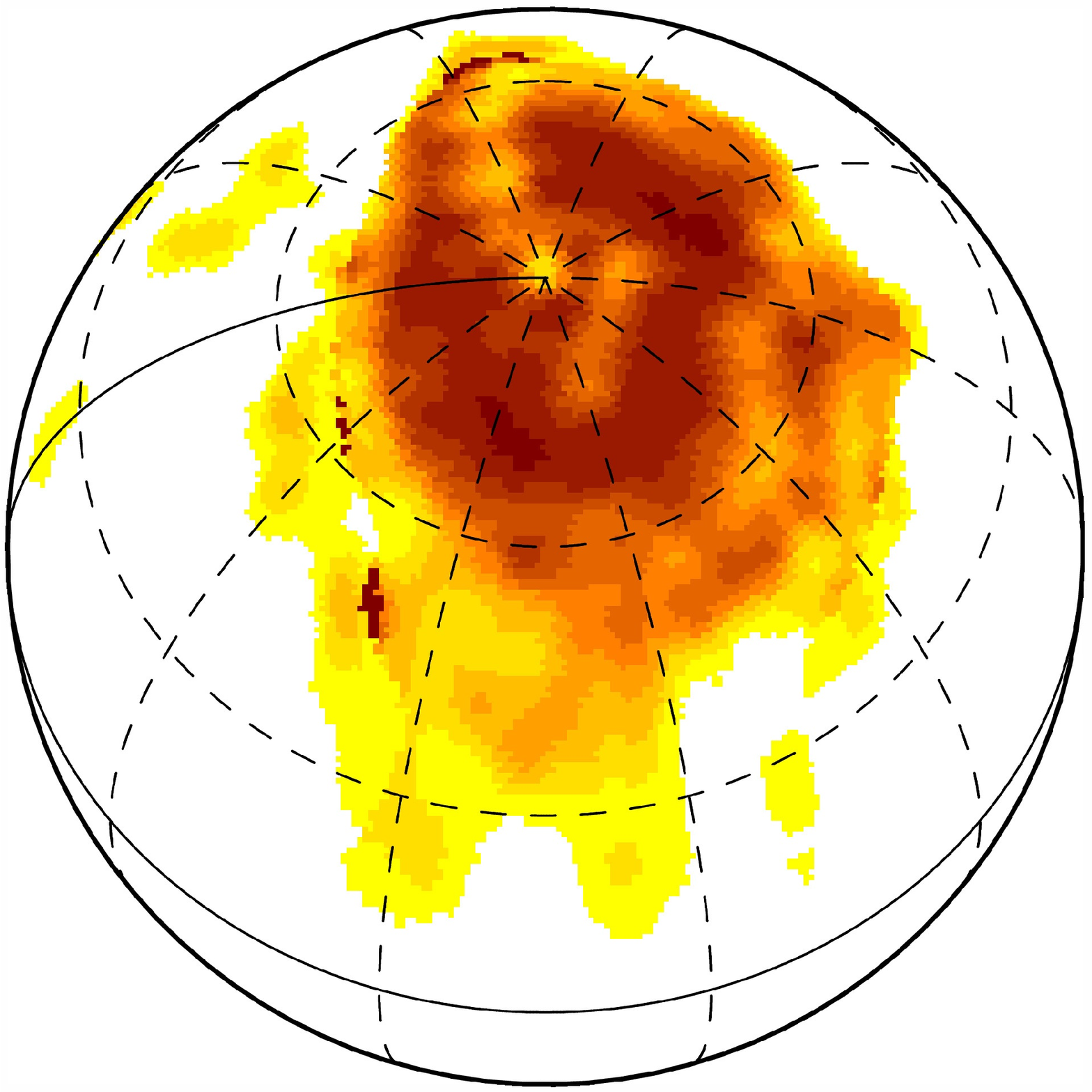}
\includegraphics[width=3.2cm]{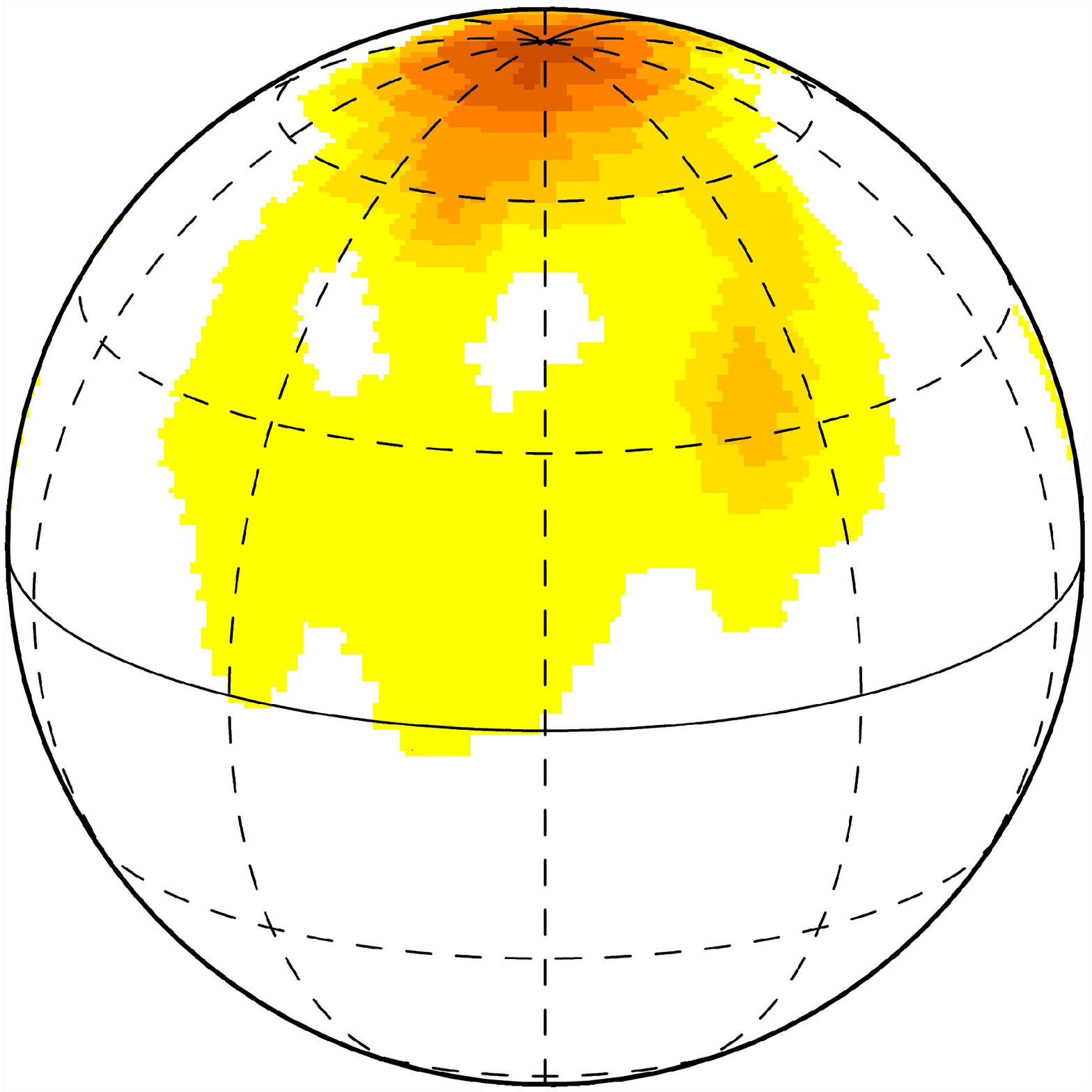}
\caption{Doppler Imaging maps for our target stars in spherical projection. Clockwise from Top Left: R58 (2003), R58 (2005), RX J0850.1-7554 (2006) and LQ Lup (2002).}
   \label{fig1}
\end{center}
\end{figure}

DI of the target stars produced reconstructed images that showed evidence of a large polar spot as well as lower latitude features on all target stars (Figure~\ref{fig1}), a result consistent with observations of other young solar type stars (i.e. \cite[Marsden et al. 2006]{MarsdenSC_2006}). 

The differential rotation of these spot features can assist in the study of the stellar dynamo, and thus differential rotation measurements were determined for our targets (see Table~\ref{tab1}). The results of these measurements are generally consistent with the results of \cite[Marsden et al. (2011)]{MarsdenSC_2011} which show an increase in differential rotation ($\delta\Omega$) with a decrease in convective zone depth for solar-type stars. In addition, both R58 and LQ Lup  show temporal evolution of their differential rotation compared to previous studies (\cite[Marsden et al. 2005]{MarsdenSC_2005} and \cite[Donati et al. 2000]{DonatiJF_2000}), a result also discovered for the active young K-dwarf AB Dor (\cite[Collier Cameron \& Donati 2002]{Collier-CameronDonati02}). These temporal changes thus suggest G-dwarfs may be winding up and down over long periods similar to lower-mass K-dwarfs, although more observations are needed to confirm such a suggestion. 

\begin{table}[h]
\begin{center}
\caption{Preliminary differential rotation ($\delta\Omega$) measurements determined for our targets. $\delta\Omega$ ranges from moderate (for R58 in 2005) to almost solid-body rotation (for RX J0850.1-7554).}
\label{tab1}
\begin{tabular}{lc}
\hline
Star (year) & $\delta\Omega$ (rad d\sups{-1}) \\
\hline
R58 (2003) & 0.109 $\pm$ 0.014 \\
R58 (2005) & 0.195 $\pm$ 0.024 \\
LQ Lup (2002) & 0.097 $\pm$ 0.004 \\
RX J0850.1-7554 (2006) & 0.004 $\pm$ 0.046 \\
\hline
\end{tabular}
\end{center}
\end{table}

In marked contrast to the moderate differential rotation of R58 and LQ Lup, the preliminary measurement for the star RX J0850.1-7554 shows a differential rotation near zero, and hence a near-solid body rotation that does not fit the trend presented in \cite[Marsden et al. (2011)]{MarsdenSC_2011}. However, due to the large errors associated with our measurement, RX J0850.1-7554 requires more observations to confirm such a low differential rotation rate.

\end{document}